\documentstyle{article}[12pt]

\def\j{{\rm i}}

\def\be{\begin{equation}}
\def\ee{\end{equation}}
\def\ben{\begin{displaymath}}
\def\een{\end{displaymath}}
\def\ba{\begin{array}{c}}
\def\ea{\end{array}}

\begin{document}

.

\vspace*{4cm}

\begin{center}
{\Large \bf
  Quasi-exact minus-quartic oscillators in strong-core regime
 }
\end{center}

\vspace{5mm}

\begin{center}
Miloslav Znojil

\vspace{5mm}

OTF, \'{U}stav jadern\'e fyziky AV \v{C}R,

250 68 \v{R}e\v{z}, Czech Republic

email: znojil@ujf.cas.cz

\end{center}

\vspace{5mm}



\section*{Abstract}

${\cal PT}-$symmetric potentials $V({x}) = -{x}^4 +\j\,B\,{x}^3 +
C\,{x}^2+\j\,D\,{x} +\j\,F/{x} +G/{x}^2$ are quasi-exactly
solvable, i.e., a specific choice of a small $G=G^{(QES)}=
integer/4$ is known to lead to wave functions $\psi^{(QES)}(x)$ in
closed form at certain charges $F=F^{(QES)}$ and energies
$E=E^{(QES)}$. The existence of an alternative, simpler and
non-numerical version of such a construction is announced here in
the new dynamical regime of very large $G^{(QES)} \to \infty$.

 \vspace*{1cm}

 \vspace{5mm}

PACS

03.65.Ca;
03.65.Ge;

\newpage

\section{Introduction}

Eight years ago Bender and Boettcher \cite{BBL} considered a
specific ${\cal PT}-$symmetric ``asymptotically repulsive"
oscillator
 \be
 H= p^2+V^{(BB)}(x),
 \ \ \ \ \ \ \
 V^{(BB)}(x)=-x^4+2\j ax^3+(a^2-2b)\,x^2+2\j \,(ab-N)\,x
 \label{QESBB}
 \ee
and conjectured and numerically verified that it possesses the
real and discrete spectrum in certain intervals of couplings $a$
and $b$ (note that while ${\cal P}$ denotes the operator of
parity, the complex conjugation ${\cal T}$ mimics time reversal so
that $a$ and $b$ must be chosen real). These authors emphasized
that their non-Hermitian model may be considered, in a way, a
``nearest neighbor" of the harmonic oscillator as it exhibits, in
a sharp contrast to its undeservedly more popular Hermitian and
asymptotically growing $+x^4$ alternative \cite{Ushveridze}, the
exceptional quasi-exact solvability (QES, \cite{Turbiner}).

By definition the latter feature means that in a way which
parallels harmonic-oscillator wave functions $\psi(\vec{x}) \sim
\exp (-\vec{x}^2/2)\times\ a \ polynomial$, {a part} of the set of
the bound states generated by the Hamiltonian~(\ref{QESBB})
remains elementary,
 \be
 \psi^{(BB)}(x) = e^{-\j\,x^3/3-a\,x^2/2-\j\,b\,x}\,\sum_{k=0}^{N}
 c_kx^k\,.
 \label{psiBB}
 \ee
This observation acquires a particular appeal in the light of the
recent increase of interest in the possible applications of
non-Hermitian models in quantum optics \cite{Geyerb} and in the
analyses of quantum chaos \cite{Geyerc} as well as in various
innovations of supersymmetric \cite{ostbe}, magnetohydrodynamical
\cite{ostbc} or particle-physics \cite{ostbd} models. During the
recent quick development of the related theory of ${\cal
PT}-$symmetric models \cite{hali} - \cite{beli} it has been,
moreover, revealed that their quantum bound states may be assigned
the standard probabilistic interpretation, provided only that one
re-defines the scalar product in Hilbert space, $\langle \cdot |
\cdot \rangle_{Dirac} \ \rightarrow \ \langle \cdot | \cdot
\rangle_{adapted}$. For this purpose one only has to introduce an
unusual, Hamiltonian-dependent metric operator $\Theta\neq I$ in a
way which proved productive in nuclear physics \cite{Geyer},
 \be
 \langle \psi_1 | \psi_2\rangle_{Dirac} \ \longrightarrow \
 \langle \psi_1 | \psi_2\rangle_{adapted}\ \equiv\
 \langle \psi_1 |\Theta| \psi_2\rangle_{Dirac}\,.
 \ee
It is now agreed \cite{ost} that the ${\cal PT}-$symmetric
Hamiltonians $H^{({\cal PT}-symmetric)}$ may be used as
phenomenological models whenever we succeed in an explicit
construction of the metric operator
$\Theta=\Theta(H)$\footnote{See, e.g., the September special issue
of Czech. J. Phys. 55 (2005), pp. 1045 - 1192 for more details.}.

The latter observations enhance the importance of the Bender's and
Boettcher's partially solvable two-parametric model (\ref{QESBB})
as well as of its straightforward three-parametric ``charged" and
``spiked" generalization, with a Coulomb and centrifugal force
added in ref.~\cite{quartic},
 \be
 V(x) = -x^4 +\j\,B\,x^3 + C\,x^2+\j\,D\,x +\j\,F/x +G/x^2\,.
 \label{five}
 \ee
Unfortunately, the phenomenological applicability of both these
models proved unexpectedly hindered by the computational
difficulties arising during the explicit construction of their
exact bound states (cf. section \ref{Maga} for a brief review). In
a reaction to such a contradictory situation we returned to this
class of models once more. We revealed and report here a
significant simplification of the QES construction which emerges
in a strongly spiked limit, i.e., for very large couplings $G\to
\infty$.

In full detail our observations will be described in
section~\ref{neg} and summarized in section~\ref{summary}
emphasizing that the new dynamical regime is complementary to the
two-parametric option~(\ref{QESBB}) of ref.~\cite{BBL} with
vanishing $G$. Our new construction may even be considered simpler
since it leads to the determination of the negative-quartic ${\cal
PT}-$symmetric QES bound states in terms of closed and compact
formulae.

\section{Quartic models and their quasi-exact solvability \label{Maga}}

\subsection{A broader family of the next-to-harmonic models?}

Among all the available exactly solvable versions of
Schr\"{o}dinger equation $H\,|\psi\rangle = E\,|\psi\rangle$ in
Quantum Mechanics, an undoubtedly exceptional position belongs to
the harmonic oscillator, the Hamiltonian of which preserves {\em
the same} differential-operator form in {\em both} the $\vec{\bf
x}-$ and $\vec{\bf p}-$representations \cite{Fluegge}. Although
such a curious ``Fourier-transformation-symmetry" property of
$H^{(HO)} = \vec{\bf p}^2+\vec{\bf x}^2$ does not survive the
transition to the ``next", quartic anharmonic oscillators, it
still may play a role in their perturbative \cite{Blankenbecler}
or continued-fraction \cite{SAO} description. Moreover, an
unexpected role of the Fourier-transformation partnership between
two {\em different} quartic oscillators has been revealed by
Buslaev and Grecchi who succeeded in proving a  strict
isospectrality between certain two ``next-to-harmonic"
quartic-oscillator models $H^{(Hermitian)}$ and $H^{({\cal
PT}-symmetric)}$ (cf.~\cite{BG}).

The subsequent increase of interest in ${\cal PT}-$symmetry in
Quantum Mechanics \cite{hali} climaxed recently, in the specific
quartic-oscillator context, with the paper \cite{Jones} where, for
a sample choice of the negative-quartic $H^{({\cal PT}-symmetric)}
\sim -x^4 $, an explicit construction of the metric $\Theta$ has
been presented as performed without {\it ad hoc} tricks and
starting simply from the first principles. The related Buslaev's
and Grecchi's results are recollected there so that, in some
sense, the circle is closed and the picture seems completed. Yet,
the description of another, viz., QES harmonic-oscillator-like
property of models $H^{({\cal PT}-symmetric)}$ deserves an
independent completion.

In a way indicated by Buslaev and Grecchi (\cite{BG}, cf. also
\cite{Classif}) and re-emphasized, e.g., by Dorey et al
\cite{Dorey}, our understanding of the various aspects of ${\cal
PT}-$symmetry may be made simpler rather than more complicated by
an introduction of the angular momentum $L$ in our ordinary
Schr\"{o}dinger equation,
 \be
 \left[-\,\frac{d^2}{dx^2} + \frac{L(L+1)}{x^2} + V(x)
 \right]\, \psi(x) =
 E \psi(x)\,.
 \label{rad}
 \ee
Traditionally one abbreviates
 \be
 L = \frac{1}{2}(d-3), \ 1+
\frac{1}{2}(d-3), \ 2+\frac{1}{2}(d-3),\
 \ldots
 \ee
in $ d \geq 3$ dimensions \cite{BG} but one may also take into
consideration the centrifugal-like spike in the potential
(\ref{five}). Thus, a generalization (\ref{QESBB}) $\to$
(\ref{five}) is to be understood as a transition to the singular
models with $F \neq 0$ and/or with
 \be
 L(L+1)+G= \ell(\ell+1)
 \neq 0,
 \ \ \ \ \ \ \ \
 \ell = \sqrt{G+\left (L+\frac{1}{2} \right
 )^2}-\frac{1}{2}\,.
 \ee
In ref.~\cite{quartic}, a theoretical merit of such a step has
been seen in the identification of the older regular model
(\ref{QESBB}) with the mere {\em special case} of
eq.~(\ref{five}). Indeed, the vanishing of the charge
$F=F^{(QES)}$ as {\em postulated} in ref.~\cite{BBL} {\em
results}, in fact, directly from the QES conditions at
$\ell(\ell+1) = 0$.

Another consequence of the formal presence of the centrifugal term
in eq.~(\ref{rad}) lies in the related possibility of a
modification of the potential (i.e., of the dynamics) by a mere
formal change of the variables in eq.~(\ref{rad})~\cite{Classif}.
This idea will not be discussed here in any detail but the
interested reader may consult ref.~\cite{decadic} for an
illustration.

\subsection{$|x| \gg 1$ asymptotics
for the decreasing quartic potentials}

The general QES recipe starting from a polynomial potential [say,
(\ref{QESBB}) or (\ref{five})] constructs its QES bound states
[exemplified here by eq.~(\ref{psiBB})] in a way described by
Magyari \cite{Magyari}. Basically, the construction parallels the
harmonic-oscillator factorization $\psi(\vec{x}) \sim \exp
(-\vec{x}^2/2)\times\ a \ polynomial$ where, for the ${\cal
PT}-$symmetric quartic model (\ref{five}) with five real
couplings, one extracts and separates the $|x| \gg 1$
asymptotically dominant part of the normalizable (i.e.,
bound-state) wave function into its exponential factor,
 \be
  \psi(x)=\exp
 \left( -\, \frac{1}{3}\,\j\,x^3 -
 \frac{1}{2} \beta \,x^2 - \j\,\gamma \,x \right)\
 \sum^{\infty}_{n=0} \omega_n\,(\j x)^{n+p}\,
 \ \ \ \ \ \ \
 \beta=B/2, \ \ \gamma=(\beta^2-C)/2\,.
 \label{anhari}
 \ee
In such a scenario and in a way extending eq.~(\ref{psiBB}) to
$\ell \neq 0$, all the QES states will be characterized by the
{\em exact} reduction of the infinite series to a polynomial,
 \be
 \omega_{N+1}=
 \omega_{N+2}= \ldots = 0\,.
 \label{termi}
 \ee
This means that our Schr\"{o}dinger eq.~(\ref{rad}) must be
integrated over a complex contour of coordinates $x \in {\cal C}$
which is bent downwards, say, towards its asymptotes
 \be
  {\cal C}_{\it le\!ft} \sim -\varrho\,e^{+\j\,\varphi_{\it le\!ft}},
  \ \ \ \ \ \ \ \ \ \ \ \ \
  {\cal C}_{\it right} \sim +\varrho\,e^{-\j\,\varphi_{\it right}},
  \ \ \ \ \ \ \ 0 < \varphi_{\it le\!ft,\,right} < \frac{\pi}{3}\,
  \ee
with the large and positive real parameter $\varrho \to \infty$.
Indeed, it is easy to verify that the exponent in
eq.~(\ref{anhari}) decreases along both these half-lines,
 \be
 e^{-\, \frac{1}{3}\,\j\,x_{\it le\!ft}^3}=
 e^{+\, \frac{1}{3}\,\j\,\varrho^3(\cos 3\varphi_{\it le\!ft}
 + \j\,\sin 3\varphi_{\it le\!ft})}
 \approx
 e^{-\, \frac{1}{3}\,\varrho^3
 \sin \varphi_{\it le\!ft}}\times {\it a\ bounded\  oscillatory\ factor}
 ,
 \ee
 \be
 e^{-\, \frac{1}{3}\,\j\,x_{\it right}^3}=
 e^{-\, \frac{1}{3}\,\j\,\varrho^3(\cos 3\varphi_{\it  right}
 - \j\,\sin 3\varphi_{\it  right})}
 \approx
 e^{-\, \frac{1}{3}\,\varrho^3
 \sin \varphi_{\it  right}}\times {\it a\ bounded\  oscillatory\
 factor}.
 \ee
As long as our problem is analytic in the whole cut complex plane
(with the cut starting in the origin and oriented upwards), the
contour ${\cal C}$ may be chosen as safely avoiding the
singularity in the origin. Thus, in terms of the effective angular
momentum $\ell$ we have to fix the sub-exponential exponent $p
\equiv -\ell$ in eq.~(\ref{anhari}) in a way compatible with both
refs.~\cite{BBL} and~\cite{quartic}.

\subsection{Magyari's QES conditions}

The insertion of the ansatz (\ref{anhari}) + (\ref{termi}) in
Schr\"{o}dinger equation (\ref{rad}) fixes the QES-compatible
value of the coupling at the linear potential term,
 \be
 D=D(N)=2(\ell+\beta\gamma-N-1)
 \ee
and imposes, furthermore, the following overcomplete linear
algebraic set of $N+2$ constraints upon the $N+1$ (arbitrarily
normalized) wave function coefficients $\omega_n$,
 \be
 (2\ell-n)(n+1)\omega_{n+1}
 +[F-2\gamma(\ell-n)]\omega_{n}
 +[E-\gamma^2+\beta(2\ell-2n+1)]\omega_{n-1}
 +2(N+2-n)\omega_{n-2}=0
 \label{emagg}
 \ee
where $n$ runs from 0 till $N-1$. With obvious abbreviations,
these equations may be re-written  as a non-square matrix problem
 \be
 \left (
 \begin{array}{ccccccc}
 S_0(F)&U_0&&&&\\
 T_1(E)&S_1(F)&U_1&&&&\\
 W_2&T_2(E)&S_2(F)&U_2&&&\\
 &W_3&T_3(E)&S_3(F)&U_3&&\\
 &&\ddots&\ddots&\ddots&\ddots&\\
 &&&W_{N-1}&T_{N-1}(E)&S_{N-1}(F)&U_{N-1}\\
 &&&&W_N&T_N(E)&S_N(F)\\
 &&&&&W_{N+1}&T_{N+1}(E)
 \ea
 \right )
 \left (
 \ba
 \omega_0\\
 \omega_1\\
 \vdots\\
 \omega_N
 \ea
 \right )=0\,.
 \label{Magus}
 \ee
It must be solved numerically in general \cite{Dubna}.

\section{Two feasible versions of the QES construction
\label{neg}}

\subsection{The domain of the small $G$, $L$ and $\ell$}

In the practical computations one may treat eq.~(\ref{Magus}) as
the two linear square-matrix eigenvalue problems
 \be
 \left (
 \begin{array}{cccccc}
 S_0(0)&U_0&&&\\
 T_1(E)&S_1(0)&U_1&&&\\
 W_2&T_2(E)&S_2(0)&U_2&&\\
  &\ddots&\ddots&\ddots&\ddots&\\
 &&W_{N-1}&T_{N-1}(E)&S_{N-1}(0)&U_{N-1}\\
 &&&W_N&T_N(E)&S_N(0)
 \ea
 \right )
 \left (
 \ba
 \omega_0\\
 \omega_1\\
 \vdots\\
 \omega_N
 \ea
 \right )=
 -F_e(E)\,
 \left (
 \ba
 \omega_0\\
 \omega_1\\
 \vdots\\
 \omega_N
 \ea
 \right )
 \ee
 \be
 \left (
 \begin{array}{cccccc}
 T_1(0)&S_1(F)&U_1&&&\\
 W_2&T_2(0)&S_2(F)&U_2&&\\
  &\ddots&\ddots&\ddots&\ddots&\\
 &&W_{N-1}&T_{N-1}(0)&S_{N-1}(F)&U_{N-1}\\
 &&&W_N&T_N(0)&S_N(F)\\
 &&&&W_{N+1}&T_{N+1}(0)
 \ea
 \right )
 \left (
 \ba
 \omega_0\\
 \omega_1\\
 \vdots\\
 \omega_N
 \ea
 \right )=
 -E_e(F)\,
 \left (
 \ba
 \omega_0\\
 \omega_1\\
 \vdots\\
 \omega_N
 \ea
 \right )
 \ee
which are complemented by the two mutual-coupling conditions
 \be
 E=E_e(F), \ \ \ \ \ \ F =
 F_e(E)\,.
 \ee
An important note is to be added, based on the inspection of
eq.~(\ref{emagg}). It reveals that in the Magyari's non-square
matrix constraint (\ref{Magus}), the upper-diagonal coefficient
$U_n= (2\ell-n)(n+1)$ can in fact vanish at $n=n(\ell)=2\ell$,
i.e., at all the half-integer effective angular momenta $\ell$. In
eq.~(\ref{rad}) this represents a constraint upon the freedom in
the choice of the spike strength $G$. Thus, whenever the integer
$2\ell$ does not exceed the dimension $N$, at least one of the
coupled secular determinants factorizes \cite{quartic} and a
significant reduction of the complexity of the algebraic QES
conditions is achieved. Still, the difficulties with the
construction grow fairly quickly with the growth of $2\ell$ even
at the integer values of this parameter.

In the light of the latter comment one may be quite surprised by
our forthcoming present main result saying that a {\em dramatic
and drastic} simplification of the recipe recurs in the asymptotic
domain where $\ell \to \infty$.

\subsection{Quartic QES models and their unexpected
duality at the large $\ell$}

A few {\em non-numerical} samples of the solution of
eqs.~(\ref{Magus}) may be found in refs.~\cite{BBL} (using
$\ell=F=G=0$ and several small $N$) and~\cite{quartic} (using
$\ell=1/2$ and $N$ up to 4, or $\ell=1$ and $N$ up to 3). Also the
results of these studies confirm that serious computational
difficulties arise and grow very quickly whenever $\ell$ grows
beyond one. In parallel \cite{Dubna}, the form of
eq.~(\ref{Magus}) appears to be perceivably simpler whenever {\em
all} the values of the {\em other} parameters $N$, $\beta$ and
$\gamma$ become negligible in comparison with the partial-wave
index $L$ and/or with the strength of the core $G$. In the latter
dynamical regime where $\ell \gg \max (N,|\beta|,|\gamma|))$ we
may omit the negligible terms from our eq.~(\ref{Magus}) and get
the leading-order version of the QES requirement,
 \be
 \left (
 \begin{array}{cccccc}
 F-2\gamma\ell&2\ell&&&\\
  E+2\beta\ell& F-2\gamma\ell&4\ell&&&\\
 2N& E+2\beta\ell& F-2\gamma\ell&6\ell&&\\
  &\ddots&\ddots&\ddots&\ddots&\\
 &&6& E+2\beta\ell& F-2\gamma\ell&2N\ell\\
 &&&4& E+2\beta\ell& F-2\gamma\ell\\
 &&&&2& E+2\beta\ell
 \ea
 \right )
 \left (
 \ba
 \omega_0\\
 \omega_1\\
 \vdots\\
 \omega_N
 \ea
 \right )=0\,.
 \label{Magusap}
 \ee
In this equation we may re-scale the coefficients
$\omega_n=h_n\ell^{-n/3}$ and subtract the leading-order
asymptotic approximants,
 \be
 F = 2\gamma\ell + 2s\ell^{2/3},
 \ \ \ \ \ \ \ \ \
 E = -2\beta\ell+2t\ell^{1/3}.
 \ee
This replaces eq.~(\ref{Magusap}) by its strictly equivalent but
strikingly simpler form
 \be
 \left (
 \begin{array}{cccccc}
 s&1&&&\\
  t& s&2&&&\\
 N& t& s&3&&\\
  &\ddots&\ddots&\ddots&\ddots&\\
 &&3& t& s&N\\
 &&&2& t&s\\
 &&&&1&t
 \ea
 \right )
 \left (
 \ba
 h_0\\
 h_1\\
 \vdots\\
 h_N
 \ea
 \right )=0\,.
 \label{Magusta}
 \ee
The phenomenologically most important and formally most remarkable
consequence of this result is that it represents another
manifestation of the Buslaev's and Grecchi's~\cite{BG} duality
between Hermitian and non-Hermitian quartic oscillators, this time
on the level of their respective QES subsets. Indeed, in the light
of ref.~\cite{jmch}, {\em the same equation} played {\em the same
role} for the {\em Hermitian} asymptotically growing potentials
 \be
  V^{(Hermitian)}({{r}}) = +{{r}}^4
 +B\,{{r}}^3 + C\,{{r}}^2+D\,{{r}} +F/{{r}}
 +G/{{r}}^2,
 \ \ \ \ \ \ \ \ \
 r \in (0,\infty)\,.
 \label{six}
 \ee
Amazingly enough, all the numerous differences between the
potentials (\ref{five}) and (\ref{six}) (the latter being defined
on the half-axis of course) disappear on the level of constraint
(\ref{Magusta}). This enables to make the rest of our present text
short. We may just cite the final (though, by the way, not so
easily derived!) results of the extensive computations as
performed in ref.~\cite{jmch}. In particular, this enables us to
summarize that the real roots $s=t=t(N)$ of eq.~(\ref{Magusta})
form the $N-$dependent and equidistant multiplets of integers,
 \be
 t(N)=t_k(N)=N-3k, \ \ \ \ \ \ k = 0, 1, \ldots, \left
 [\frac{N}{2}\right ]\,.
 \ee
This means that the physically acceptable solutions of our present
${\cal PT}-$symmetric  $\ell \gg 1$ QES problem exist and occur in
the multiplets with the following asymptotic energies and charges,
 \be
 E = -2\beta\ell+2(N-3k)\ell^{1/3}+\ldots,
 \ \ \ \ \ \
 F = 2\gamma\ell + 2(N-3k)\ell^{2/3}+\ldots
, \ \ \ \ \ k = 0, 1, \ldots, \left
 [\frac{N}{2}\right ]\,.
 \ee
One may now return to the elementary recurrences (\ref{Magusta})
and evaluate,  very easily, the coefficients $\omega_n$ of the
wave functions in the same next-to-leading order approximation. In
the light of the existing thorough analysis of this problem in the
dual Hermitian context \cite{jmch}, this task may already be left
to the readers as an exercise.

\section{Summary \label{summary}}

Could we view the quasi-exactly solvable ${\cal PT}-$symmetric
quartic potentials as a choice, in some sense, ``next" to the
popular harmonic oscillator? In our paper we tried to support an
affirmative answer.

During our study we felt particularly motivated by the technical
difficulties arising in connection with the explicit construction
of the quartic QES charges. Although, implicitly, they are defined
by the coupled pair of the Magyari's polynomial algebraic
equations for two unknowns, their practical determination must
usually rely upon the computerized, Gr\"{o}bner-basis-based
algebraic manipulation techniques and numerical root-searching
\cite{Groebner}. In addition, it is quite unpleasant that the
complexity of the latter algorithm grows fairly quickly with the
growth of the degree $N$ of the polynomial wave functions as well
as with the growth of the angular momentum $\ell$. Finally, the
situation significantly worsens whenever $\ell$ ceases to be a
half-integer \cite{quartic}.

We were encouraged by the well known fact that, quite often, the
dependence of bound state on the angular momentum may get
simplified in an asymptotic regime \cite{Omar}. In the latter
direction, our attempt proved successful. We found that several
large$-\ell$ properties of our non-Hermitian model (like, e.g.,
the subtle QES-related cancellations of the separate elements in
the infinite power series in $x$) find in fact quite close
parallels in its self-adjoint predecessors. Many differences
(e.g., the occurrence of complex couplings or the deformability of
the integration contours in ${\cal PT}-$symmetric case) proved
inessential. We arrived at the final version (\ref{Magusta}) of
the Magyari's equation which is, from the formal point of view,
identical with the equations encountered in Hermitian cases (so,
we could also employ their well known solutions in our
construction).

In conclusion we may add that whenever necessary, one may leave
the asymptotic domain and switch attention to the finite effective
angular momenta $\ell \ll \infty$. The necessary mathematics may
be found in the modified Rayleigh-Schr\"{o}dinger perturbation
recipe as adapted to non-square matrices in ref. \cite{pert}. It
is worth emphasizing that it makes the full use of the
finite-dimensional character of the Magyari's re-formulation of
our present negative-quartic QES Schr\"{o}dinger equation. For
this reason it may be recommended as an efficient and systematic
source of higher-order corrections.

\section*{Acknowledgment}
%

The research was partially supported by the projects LC06002 and
IRP AV0Z10480505.

\newpage

\end{document}